\documentclass[preprint,12pt]{elsarticle}

\usepackage{graphicx}
\usepackage{epstopdf}
\usepackage{tikz}

\usepackage{amssymb}
\usepackage{color}

\newcommand{\bu}{\textbf{u}}

\journal{Journal of Computational Physics}

\begin{document}

\begin{frontmatter}

\title{A multiple--resolution strategy for Direct Numerical Simulation of scalar turbulence }

\author{R. Ostilla--Monico$^1$, Yantao Yang$^1$, E. P. van der Poel$^1$, D. Lohse$^1$ and R. Verzicco$^{1,2}$}

\address{$^1$ Physics of Fluids, MESA+ institute, University of Twente, Enschede, The Netherlands, \\
$^2$ Department of Industrial Engineering, Universit\`{a} di Roma ``Tor Vergata,''   Roma, Italy. }

\begin{abstract}
In this paper a numerical procedure to simulate low diffusivity scalar turbulence is presented.
The method consists of using a grid for the advected scalar with a higher spatial resolutions than that of the momentum.
The latter usually requires a less refined mesh and integrating both fields on a single
grid tailored to the most demanding variable, produces an unnecessary computational overhead.
A multiple resolution approach is used also in the time integration in order to maintain the
stability of the scalars on the finer grid.
The method is the more advantageous the less diffusive the scalar is with respect to momentum, therefore it is particularly well suited for large Prandtl or Schmidt number flows. However, even in the case of equal diffusivities the present procedure gives CPU time and
memory occupation savings. The reason is that the absence of the pressure term in the scalar equation leads to much steeper gradients in the scalar field as compared to the velocity field.
\end{abstract}


\end{frontmatter}

\section{Introduction}
\label{sec:intro}

Countless phenomena in Nature and technology involve one 
or more scalar fields that are advected and diffused by a turbulent 
flow.  The dilution of pollution in the atmosphere \cite{arya}, the transport of nutrients in oceans 
\cite{pel03}, the cooling or heating of devices \cite{incro}
and the buoyancy--driven currents generated by natural-- \cite{sig94,lin99} and 
double--diffusive  \cite{tur85,scm94} convection are just few examples among many. 

Numerical simulations have proven to be very helpful in unravelling the 
complex physics behind these phenomena \cite{ahl09} even if the calculations have shown 
to be more demanding than expected, taking up to millions of CPU hours in
recent studies \cite{ste11}.

The common understanding of the problem is that in three--dimensional turbulent
flows, there is a cascade from the largest
towards the small spatial scales up to a lower limit that is determined by the
diffusivity. As each field has its own diffusivity, these scales can have
different magnitudes. In direct numerical simulation (DNS) the mesh size must be  
smaller than the smallest among them: This requirement quickly renders DNS
infeasible. Denoting as $\eta_K$ the smallest (Kolmogorov) scale of the 
momentum field, we can calculate the analogous quantity for a scalar field $S$ as $\eta_B
= \eta_K/Sc^{1/2}$, also called the Batchelor scale, with $Sc = \nu/\kappa_S$ the
Schmidt number defined as the ratio of the kinematic viscosity $\nu$ and the 
scalar diffusivity $\kappa_S$, respectively. 
In some cases, like sugar in water, the Schmidt number exceeds $10^3$ resulting in a  Batchelor scale of $\eta_B \simeq \eta_K/30$. With equal grid resolutions for the scalar and the momentum fields, this entails that the momentum field is overresolved by a factor of approximately $30$ in each direction.
The problem is exacerbated by the fact that a scalar is described by only a single quantity, while momentum is a vector field satisfying the incompressibility condition or other related constraints. 
This implies
that the solution of the momentum alone generally takes an order of $90\%$ of the total
CPU time of a simulation and therefore resolving it on an unnecessary fine mesh is not desirable.

\begin{figure}[htb!]
\begin{center}
 \includegraphics[width=0.95\textwidth]{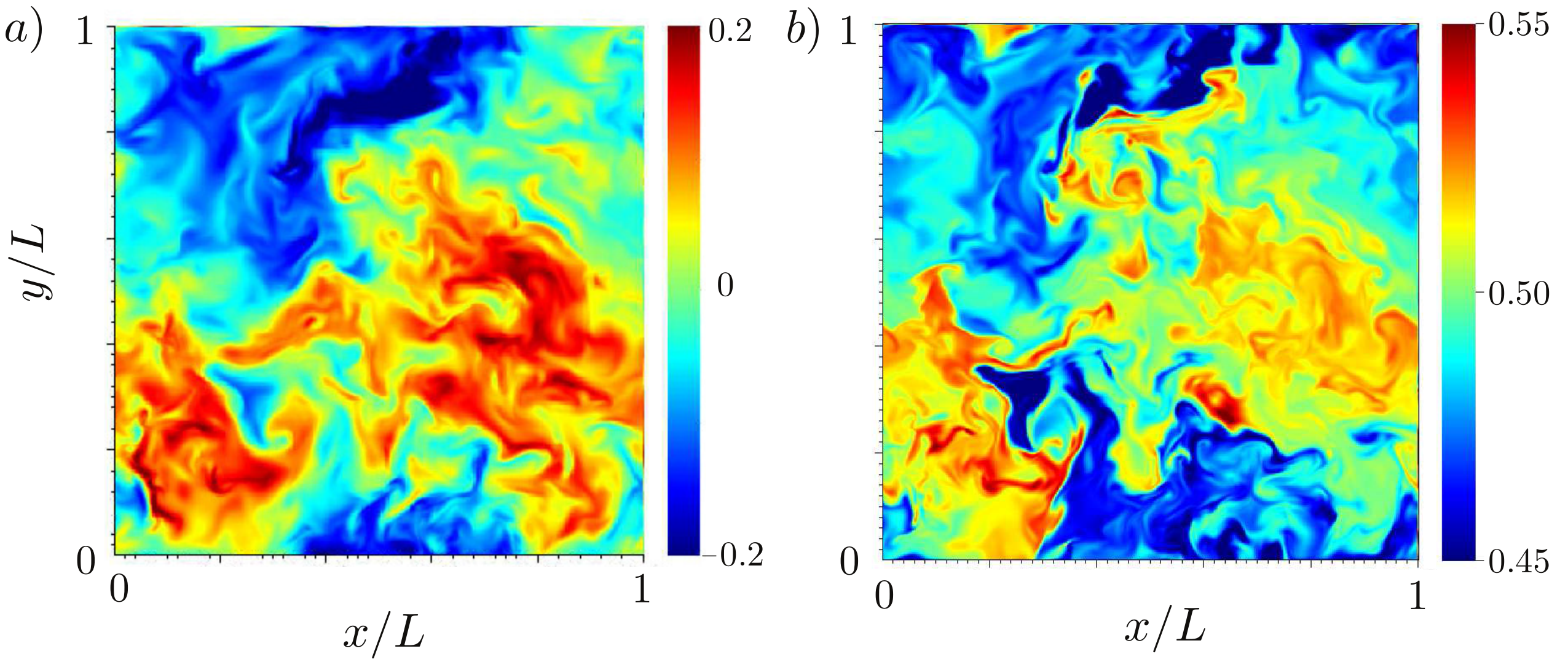}
\end{center}
\caption{A horizontal plane halfway between the plates for a Rayleigh-B\'enard 
simulation in a Cartesian geometry at $Ra=10^{10}$ and
Prandtl number $Pr=1$. {\it a)} vertical velocity, red indicating rising fluid
while blue indicates falling fluid, {\it b)} temperature, red indicating
hot fluid and blue indicating cold fluid. Even though the Prandtl number
is one, much sharper gradients can be seen in the right panel. }
\label{fig:introvis}
\end{figure}

The above scenario essentially derived from dimensional analysis, however, does not give the complete picture since it does not account for
the structure of the equations. In fact, the na\"ive comparison between the Kolmogorov and Batchelor scales suggests that for $Sc\approx 1$, $\eta_K \simeq \eta_B$ although in practice the resolution requirements for the momentum and the scalar fields are not the same. Visual evidence of the latter statement can be obtained from the instantaneous snapshots of figure \ref{fig:introvis} showing horizontal cross-sections of temperature and vertical velocity in a thermally driven turbulent flow.  In this flow, the fluid hotter than the average temperature ($0.5$ in nondimensional units) generates upward buoyancy and therefore positive vertical velocity (and vice versa). Although the two fields are very well correlated on the large scales, the sharp fronts of the scalar field do not have an analogous counterpart in the momentum distribution and this results in a different resolution requirement for scalar and momentum.  We will see that the main reason for this difference is the presence of the pressure term in the momentum equation that makes the dynamics non-local and tends to smooth the intense steep fronts.
A problem related to the non-local equations for 
momentum is their high
computational cost and the detrimental implications on the parallel performance. Therefore, the possibility of using different meshes for momentum and scalars opens the door
to very large gains in performance, not only by reducing the amount of operations,
but also the communication of data among processors and the total memory usage. This consideration motivates the present paper that describes a 
strategy for efficiently simulating scalar driven turbulent flows with 
different spatial resolutions for momentum and 
scalar fields. 

In this study,
the multiple resolution strategy will be mainly applied to Rayleigh-B\'enard (RB) convection,
the flow of a fluid vertically confined by a top cold and a bottom hot plate. 
RB is a particularly suitable example for the present application since the flow is
driven by the temperature (scalar) field whose diffusivity can be changed by the 
Prandtl number $Pr$. In addition, analytical exact relations are available for this problem
that can be used to check the correctness of the numerical results. It is worth mentioning that in RB convection the forcing comes from the
heated and cooled surfaces where viscous and thermal boundary layers develop. Since they become	
thinner as the forcing strengthens, the resolution requirements in these boundary layers become more stringent than in the bulk. An extensive
analysis of the problem can be found in Ref. \cite{shi10} where all the details,
estimates and guidelines for numerical simulations are given. Here, it suffices to mention that
a non-uniform mesh is required in the wall normal direction such to cluster the nodes
within the boundary layers. Nevertheless, even if the grid spacing at the wall is much finer
than that in the bulk, the volume of fluid within these layers is at most a few percent
of the total and the nodes allocated there are only a small fraction of the whole mesh.

We will additionally show another numerical example. Namely, double diffusive convection (DDC),
in which the flow is driven by two scalars with very different diffusivities
and with opposite, stabilizing and destabilizing, effects on the flow. In
this case the multiple resolution strategy is even more advantageous and it
allows for the simulation of flow regimes that otherwise would be out of reach.

The organization of the paper is the following.
In the next section we describe the governing equations and the numerical method.
Section \ref{sec:sres} quantifies the differences of momentum and scalar gradients
and presents some analytical
exact relations for RB. The section closes with the results of reference
simulations obtained on a standard single grid.
In Section \ref{sec:mres} the multiple resolution strategy and numerical details 
are explained. Finally, Section \ref{sec:res} discusses the results and the computational performance
of the method for RB flow and double diffusive convection.
Closing remarks are reported in section \ref{sec:conc}.

\section{Governing equations and numerical method}
\label{sec:eq}

The incompressible Navier--Stokes equations with the Boussinesq approximation for thermal convection,
in nondimensional form, read:

\begin{equation}
 \nabla\cdot\bu = 0,
 \label{eq:continuity}
\end{equation}

\begin{equation}
 \displaystyle\frac{\partial \bu}{\partial t} + \bu\cdot\nabla\bu = -\partial_i p + \displaystyle\sqrt{\frac{Pr}{Ra}} \nabla^2 \bu + \theta \textbf{e}_z,
 \label{eq:navierstokes}
\end{equation}

\begin{equation}
 \displaystyle\frac{\partial \theta}{\partial t} + \bu\cdot\nabla\theta = \displaystyle\sqrt{\frac{1}{RaPr}} \nabla^2 \theta,
 \label{eq:scalareq}
\end{equation}

\noindent where $\bu$ is the velocity, $p$ the pressure, $\theta$ the temperature 
(rescaled such that it is one at the hot plate and zero at the cold plate) and $\textbf{e}_z$ the unitary vector in the anti-parallel direction to gravity, which is also the plate-normal direction. The Rayleigh number $Ra$ is the
non-dimensional temperature difference 
defined as $Ra=g\beta_T(T_h-T_c) L^3/\kappa_T\nu$, where $\nu$ is the kinematic viscosity, 
$\beta_T$ the isobaric thermal expansion coefficient and $\kappa_T$ the thermal diffusivity of the fluid,
$g$ is the gravity,
$L$ the distance and $T_h$ and $T_c$ the temperature at the hot and cold plates, respectively. The Prandtl number $Pr$, 
which is the temperature analogue of the Schmidt number, is defined as $Pr=\nu/\kappa_T$.

The integration of the above system is performed by a fractional timestepping \cite{kim85} with the 
modifications proposed in Ref. \cite{ver96}. In short, a provisional velocity $\hat{u_i}$ is computed 
from the previous field $u_i^n$ using the old pressure $p^n$
\begin{equation}
\frac{\hat{u_i} -u_i^n}{\Delta t} = -\partial_i p^n -N_i^{n+1/2} + D_i^{n+1/2},
\end{equation}
\noindent $N_i^{n+1/2}$ contains the non-linear terms and the temperature forcing while $D_i^{n+1/2}$ has
the viscous terms: The former are computed explicitly in time, the latter implicitly.
The flow incompressibility is then enforced by a pressure correction 
that takes the form of a Poisson equation $\nabla^2 \phi=\partial_i \hat{u_i}$ whose solution is the most computationally demanding step, especially on non--uniform grids.  In addition, the Poisson equation is non--local and this has consequences on code parallelization, requiring the largest amount of communication. Once the scalar $\phi$ is obtained, the velocity $\hat{u_i}$ is projected onto the solenoidal field $u_i^{n+1}$ and the new pressure $p^{n+1}$ can be computed.
The advancement of the temperature is performed directly through
\begin{equation}
\frac{\theta^{n+1} -\theta^n}{\Delta t} = -M^{n+1/2} + V^{n+1/2},
\end{equation}
\noindent where, again, $M^{n+1/2}$ contains the explicit nonlinear terms and $V^{n+1/2}$ the
implicit diffusive terms.

All the variables are discretized by central second--order finite--differences on a staggered grid
and the time advancement of the solution is obtained by a low--storage third--order Runge--Kutta
scheme. Further details can be found in Ref. \cite{ver96}.

\section{Pressure effects and heat transfer in RB}
\label{sec:sres}

\begin{figure}[htb!]
\begin{center}
 \includegraphics[width=0.95\textwidth,clip]{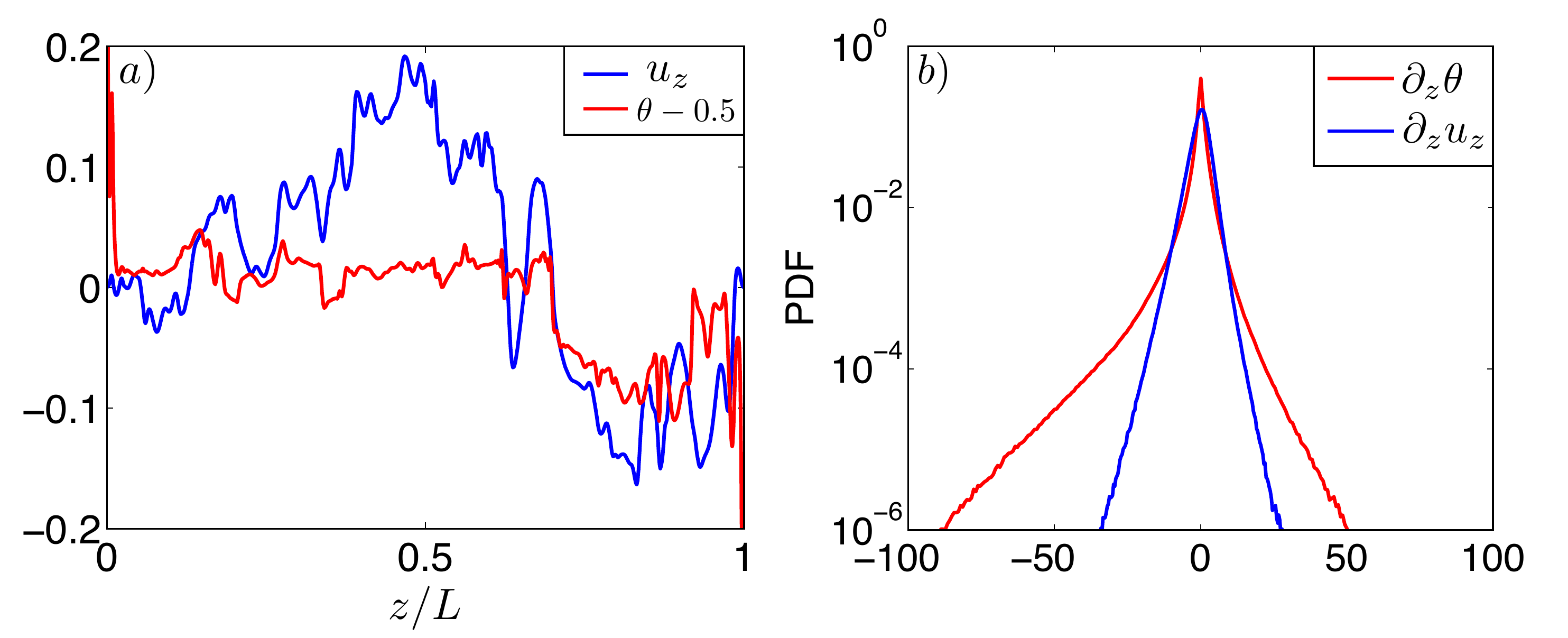}
\end{center}
\caption{{\it a)} Instantaneous $\theta$ and $u_z$ profiles as a function of the vertical coordinate $z/L$ for $Ra=10^{10}$, 
$Pr=1$ RB simulation. {\it b)} PDF of $\partial_z\theta$ and $\partial_zu_z$ for the same simulation in the bulk.}
\label{fig:thetaq3pdfs}
\end{figure}

As mentioned in the introduction,
the Prandtl number gives the ratio of momentum 
to thermal diffusivity, and even if $Pr$ is of order unity,
temperature and momentum do not have the same gradient magnitudes (cf.~figure \ref{fig:introvis}).
We quantify this statement in figure \ref{fig:thetaq3pdfs}({\it a}) by showing 
instantaneous temperature $\theta$ and vertical velocity $u_z$ profiles across a vertical line
from a RB simulation:
Much steeper gradients, almost `shock' like,
can be seen in the temperature field. These steep gradients are
smoothed in the vertical velocity, owing to the pressure effects and
this lowers the resolution requirements of momentum with respect to temperature.
This observation is further corroborated by figure \ref{fig:thetaq3pdfs}({\it b})
showing the probability density functions
of $\partial_z\theta$ and $\partial_zu_z$ computed in the bulk of the flow without
the boundary layers. Extreme gradients can be seen to be more likely for $\theta$ thus evidencing
a more intermittent behaviour. This behaviour has been extensively studied, and it is a well established fact that the intermittency corrections to the structure function exponents are much larger for scalars than for velocity \cite{shr00,loh10}. Indeed, these shock-like fronts become much sharper with increasing Reynolds number, and thus increasing small-scale intermittency \cite{sre91,pum94,pum98,cel00,war00,moi01}.

We have argued
that the absence of the pressure term in the scalar equation is the reason for
the localized steep gradients and this idea can be confirmed by the Burgers equation:
\begin{equation}
\partial_t u + u \partial_x u = \nu_B \partial_{xx} u,
\label{eq:burg}
\end{equation}
which is often used to test numerical schemes since it is a simple one--dimensional
partial differential equation that still retains the essential features of the
more complex Navier--Stokes equations: The unsteady term, a quadratic non--linear term
and a strictly dissipative viscous term \cite{bur48}.

A one--to--one comparison with equation (\ref{eq:navierstokes}), 
however, evidences the absence of a pressure
term and this causes the solution $u(x,t)$ to develop `shock'--like discontinuities in a finite time
and for finite values of the diffusivity $\nu_B$.

\begin{figure}[htb!]
\begin{center}
 \includegraphics[width=0.97\textwidth]{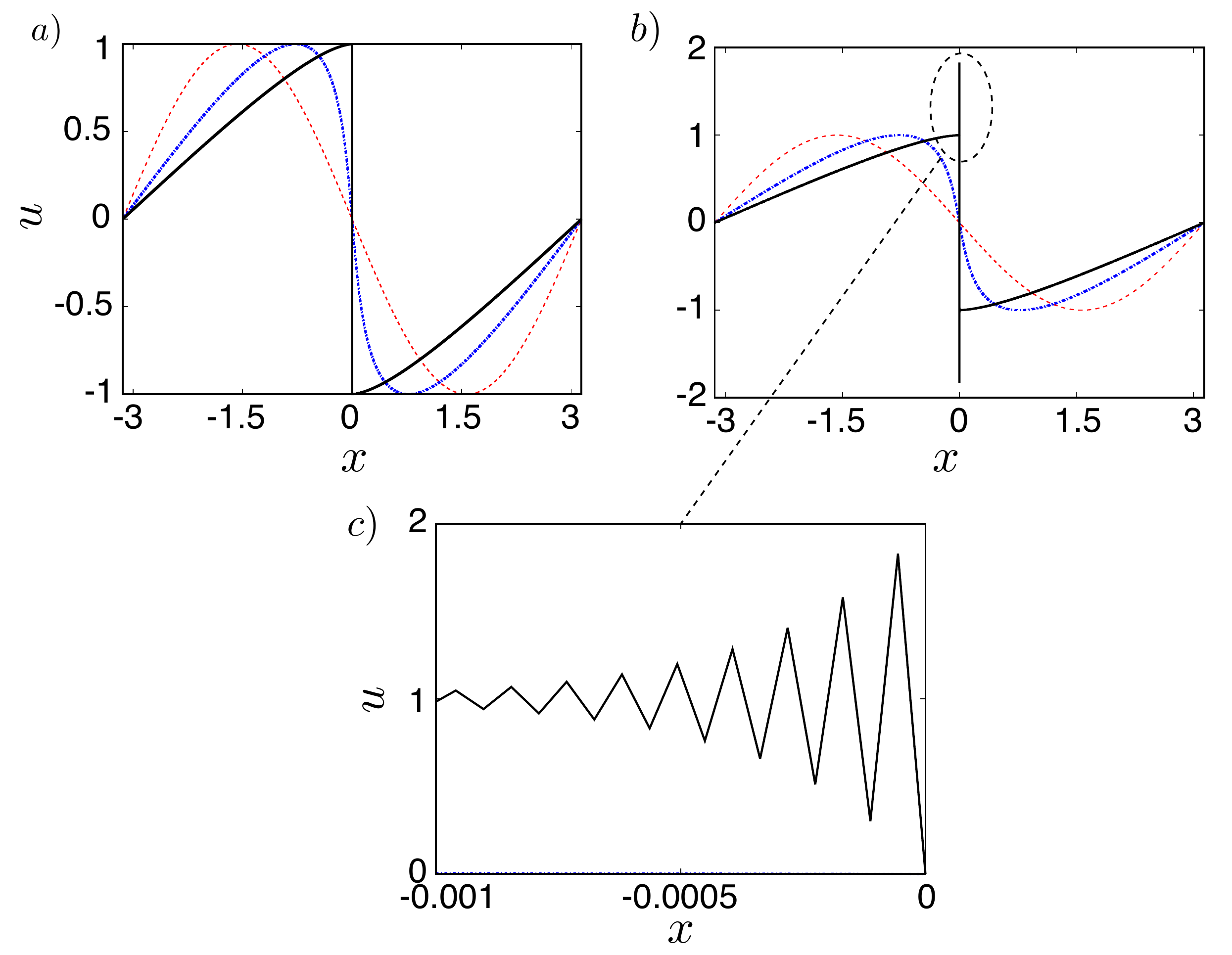}
\end{center}
\caption{Spatial profiles at various times of the solution of the Burgers equation for $\nu_B = 5\times 10^{-6}$:
 dashed $t=0$, dotted $t=0.2$, chaindot $t=0.4$ and solid $t=0.8$.
 {\it a)} resolved simulation with $\overline{\Delta x} = 10^{-4}$, 
 {\it b)} underresolved simulation with $\overline{\Delta x} = 4\times 10^{-4}$, 
 {\it c)} enlargement of {\it b)} in the region of the steep gradient.
}
\label{fig:burg}
\end{figure}

\begin{figure}[htb!]
\begin{center}
 \includegraphics[width=0.95\textwidth]{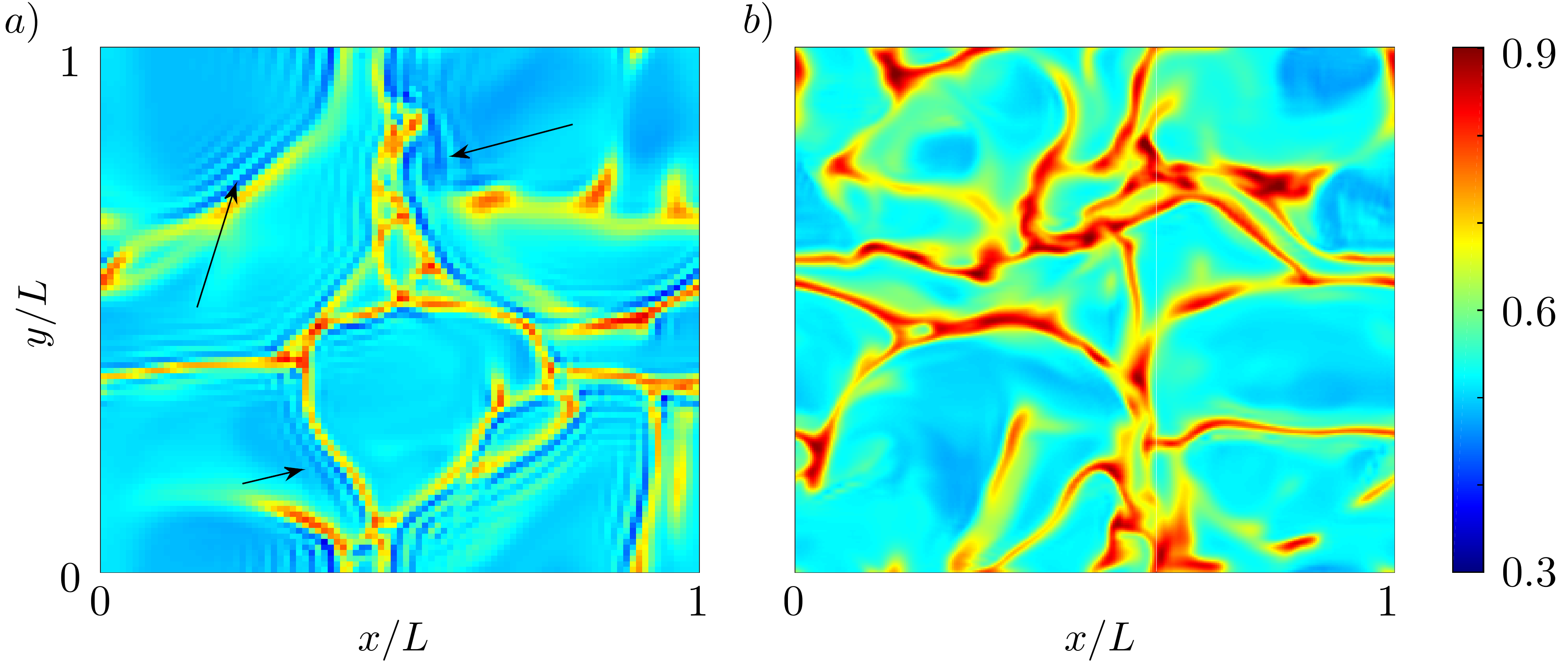}
\end{center}
\caption{ \emph{a)} Snapshot of an underresolved flow field for $Ra=10^9$ and $Pr=1$. The characteristic ``wiggles'' of the underresolved temperature field can be clearly appreciated (see arrows). \emph{b)} Snapshot of a properly resolved velocity field.}
\label{fig:underresolved}
\end{figure}

As an example, in figure \ref{fig:burg}, a solution obtained for an initial condition $u(x,0) = \sin x$ in the
domain $ -\pi \leq x < \pi$ and $\nu_B = 5\times 10^{-6}$ is shown. The gradient at $x=0$ becomes so steep
that a fine nonuniform mesh clustered at the centre of the domain is necessary to properly capture the
solution (figure \ref{fig:burg}(\emph{a}). 
For under-resolved meshes (figure \ref{fig:burg}(\emph{b}) the solution unphysically overshoots 
the initial extrema $u_{min} = -1$ and $u_{max} = 1$
and spurious oscillations are generated in the region of the steep gradient 
(figure \ref{fig:burg}(\emph{c}) resembling the behaviour of
underresolved scalar fields.
A very similar phenomenon is observed in a simulation with an underresolved temperature 
field as shown in figure \ref{fig:underresolved}.

In RB flow, one of the interesting quantities is the heat flux $Q$ transferred from one
plate to the other. In non-dimensional form this is the Nusselt number, $Nu=Q/\kappa_T(T_h-T_c) L^{-1}$.
This is not only interesting from a physical point of view, but also as a monitoring variable since it has been observed
\cite{ste10} that when the separately calculated Nusselt number converge and are grid independent, at least all quantities up to second order statistics are properly resolved.

There are several ways to calculate $Nu$, either by measuring the convective
heat transport in the system 
\begin{equation}
Nu(z) =\sqrt{RaPr} \langle u_z\theta \rangle_{A,t}+1, 
\label{def1}
\end{equation}
or by using the exact relationships derived from global balances \cite{shr90} of kinetic energy,
\begin{equation}
 \epsilon_\nu = \nu U_f^2 L^{-2} \langle [\partial_i u_j]^2 \rangle_{V,t} = \nu^3L^{-4}(Nu-1)RaPr^{-2},
\label{def2}
\end{equation}
and thermal energy,
\begin{equation}
 \epsilon_\theta = \kappa_T (T_h-T_c)^2 L^{-2} \langle [\partial_i \theta]^2 \rangle_{V,t} = \kappa_T(T_h-T_c)^2L^{-2}Nu.
\label{def3}
\end{equation}
Here the subscripts $t$, $A$ and $V$ denote, respectively, averages in time, horizontal homogeneous planes 
and the whole fluid volume, and $U_f$ the free-fall velocity $U_f=\sqrt{\beta g(T_h-T_c)L }$. 
Although equation (\ref{def1}) depends on $z$, once
the equilibrium is reached its value becomes constant. This condition is
requested to assess the statistical convergence of the results.

From here on, we denote $Nu$ with a subscript, either
$Nu_{u_z\theta}$, $Nu_{\epsilon_\nu}$ or $Nu_{\epsilon_\theta}$ 
which specifies the particular equation, i.e. (\ref{def1})--(\ref{def3}) respectively, with which $Nu$ is calculated. In addition, we also denote the Nusselt number calculated by the temperature gradient at the wall as $Nu_{\theta_w}$.
All definitions are equivalent analytically, but they involve gradients or square
gradients of the variables that, when calculated numerically, can deviate from
each other if the
simulations do not have enough spatial resolution to capture 
the smallest flow scales:
Their comparison can thus be used as a test for the adequacy of the mesh \cite{ste10}. 

\begin{figure}[htb!]
\begin{center}
 \includegraphics[width=0.98\textwidth]{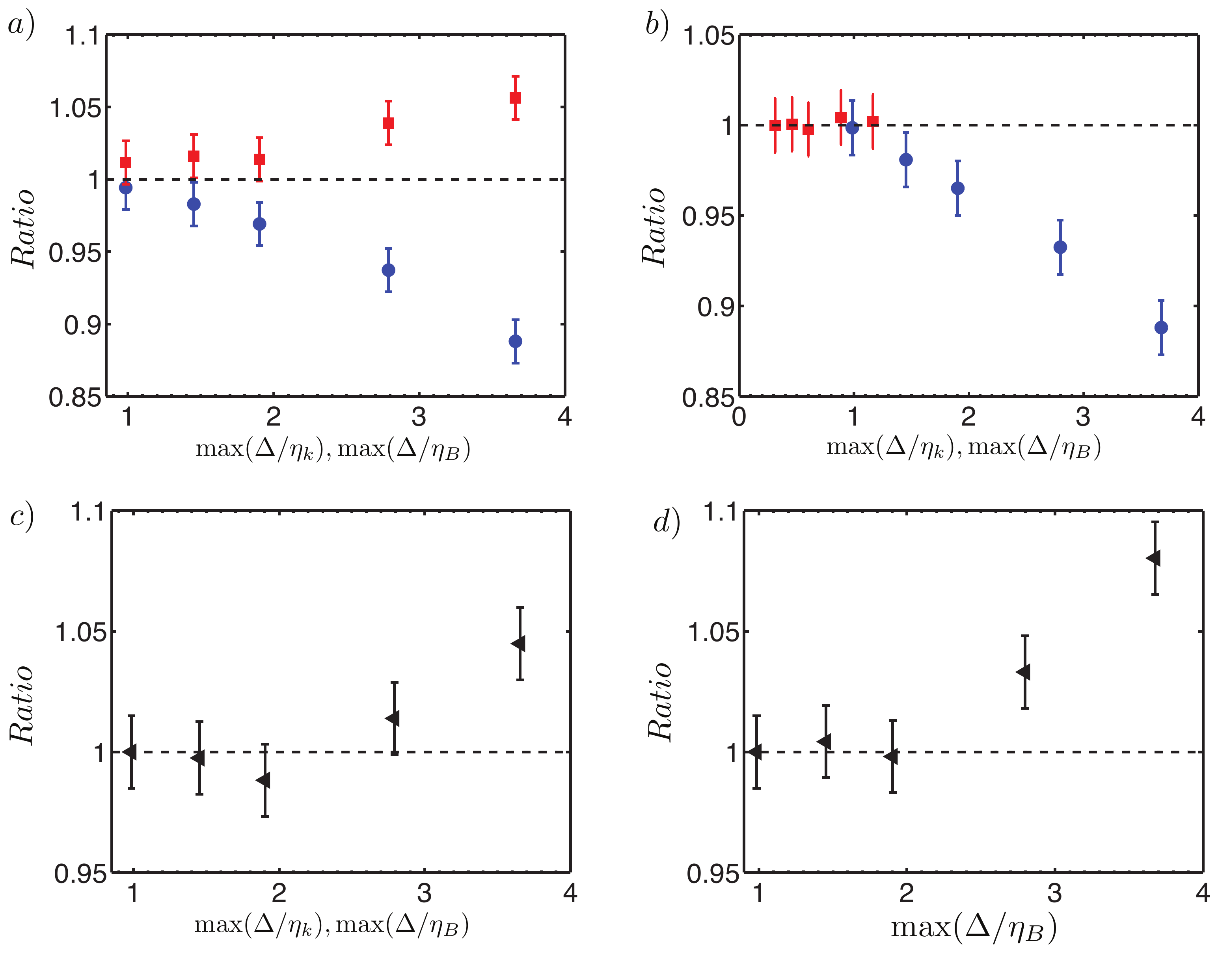}
\end{center}
\caption{({\it a}) ratio between different ways of calculating $Nu$ against grid size for a rectangular RB simulation
with $Ra=10^9$ and $Pr=1$. Red squares are $Nu_{\epsilon_\nu}/Nu_{u_z\theta}$, blue circles are $Nu_{\epsilon_\theta}/Nu_{u_z\theta}$.  ({\it b)} same as ({\it a)} for $Pr=10$. $Nu_{\epsilon_\theta}$ is plotted against
$\max(\Delta/\eta_B)$ while $Nu_{\epsilon_\nu}$ is plotted against $\max(\Delta/\eta_B)$. For $Pr=1$, $\eta_B\equiv\eta_K$.
{\it c)} convergence to an asymptotic value of 
$Nu_{u_z\theta}$ for $Ra=10^9$ and $Pr=1$. ({\it d)}: same as ({\it c)} for $Pr=10$.}
\label{fig:nussingleres}
\end{figure}

Figures \ref{fig:nussingleres}({\it a}) and \ref{fig:nussingleres}({\it b}), show the ratios 
$Nu_{\epsilon_\theta}/Nu_{u_z\theta}$ and $Nu_{\epsilon_\nu}/Nu_{u_z\theta}$
for RB simulations performed on the same grid for momentum and temperature at
$Ra=10^9$ and $Pr=1$ or $Pr=10$.
Resolutions between $96^2\times192$ and $384^2\times768$ were used, the 
larger number corresponding to the vertical (wall--bounded) direction. An aspect ratio of $\Gamma=1$ was used in both directions, meaning that the computational domain is cubic. Points were clustered near the wall using a Chebychev--like distribution according to the prescriptions of Ref.\ \cite{shi10}. The Kolmogorov scale is computed from the kinetic energy dissipation rate using $\eta_K/L=(\nu^{3}/\epsilon_\nu)^{1/4}$ and the Batchelor scale $\eta_B/L = \eta_K/L Pr^{-1/2}$.
As mentioned above, the various expressions for $Nu$ should be equivalent, their ratios however, approach the unity limit only when the normalized mesh sizes $\Delta/\eta_K$ and $\Delta/\eta_B$ decrease and they do not converge at the same rate. In particular it can be noted that $Nu_{\epsilon_\nu}/Nu_{u_z\theta}$ tends to unity for larger grid spacings than $Nu_{\epsilon_\theta}/Nu_{u_z\theta}$ even for $Pr=1$ and this corroborates our hypothesis  that a finer resolution is needed for the scalar than for momentum. Using an identical mesh to spatially discretize both momentum and the scalars therefore produces an overhead in computational resources that is redundant.

At the highest resolution, all the definitions converge to the same value (within an
uncertainty of $2$--$3\%$), therefore we will refer to it as $Nu_{ref}$ without specifing 
the particular expression and figure \ref{fig:nussingleres}({\it c}) and \ref{fig:nussingleres}({\it d}) show the convergence 
of $Nu_{u_z\theta}$ to this asymptotic value.

\section{The multiple resolutions strategy}
\label{sec:mres}

\subsection{Multiple resolution strategy in space}
\label{sec:mress}

\begin{figure}
\centering
\hspace{1cm}
\begin{tikzpicture}
\node [above left,black] at (0,3) {\emph{a})};
\draw (0,0) rectangle (3.5,3);
\filldraw[black] (1.75,1.5) circle(1mm);
\node [above right,black] at (1.75,1.5) {$S,p$};
\draw [->,thick] (0,1.5) -- (0.6,1.5);
\node [above right,black] at (0,1.5) {$u$};
\draw [->,thick] (3.5,1.5) -- (4.1,1.5);
\node [above right,black] at (3.5,1.5) {$u$};
\draw [->,thick] (1.75,0) -- (1.75,0.6);
\node [above right,black] at (1.75,0) {$v$};
\draw [->,thick] (1.75,3) -- (1.75,3.6);
\node [above right,black] at (1.75,3) {$v$};

\node [above left,black] at (5,3) {\emph{b})};
\draw (5,0) rectangle (11,3);
\draw [thin, dashed] (7.5,0) -- (7.5,3);
\draw [thin, dashed] (5,1.5) -- (11,1.5);
\filldraw[black] (6.25,0.75) circle(1mm);
\node [above right,black] at (6.25,0.75) {$S$};
\filldraw[black] (9.25,0.75) circle(1mm);
\node [above right,black] at (9.25,0.75) {$S$};
\filldraw[black] (6.25,2.25) circle(1mm);
\node [above right,black] at (6.25,2.25) {$S$};
\filldraw[black] (9.25,2.25) circle(1mm);
\node [above right,black] at (9.25,2.25) {$S$};
\filldraw[black] (8,1.5) circle(1mm);
\node [above right,black] at (8,1.5) {$p$};
\draw [->,thick] (5,1.5) -- (5.6,1.5);
\node [above right,black] at (5,1.5) {$u$};
\draw [->,thick] (11,1.5) -- (11.6,1.5);
\node [above right,black] at (11,1.5) {$u$};
\draw [->,thick] (8,0) -- (8,0.6);
\node [above right,black] at (8,0) {$v$};
\draw [->,thick] (8,3) -- (8,3.6);
\node [above right,black] at (8,3) {$v$};

\end{tikzpicture}
\caption{Location of pressure, temperature and velocities of a 2D simulation cell. The third dimension ($z$) is omitted for clarity. As on an ordinary staggered scheme, the velocity vectors are placed on the borders of the cell and pressure is placed in the cell center. The temperature is placed on the same nodes as the vertical velocity, to ensure exact energy conservation.}
\label{fig:discr}
\end{figure}
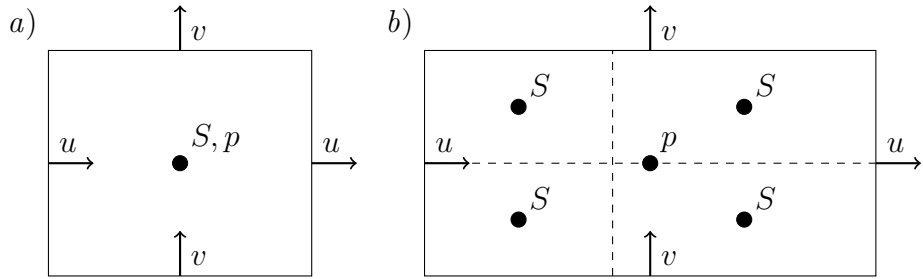

In this subsection we present a method to decouple the spatial discretization of the scalars and the momentum, which allows for  large computational savings. This is achieved by refining every cell of a base mesh $\mathcal{M}^i$ times in each $i$-th direction. A simplified two--dimensional sketch of this procedure is shown in figure \ref{fig:discr}. On the left, the location of the scalars and velocities in the standard single mesh is shown for a staggered arrangement. There, the velocity components are at the centres of the cell faces, while the pressure and the scalar are discretized at the centre of the cell volume. The right panel shows a case with velocity and pressure on the base grid, and a doubly refined ($\mathcal{M}^x=\mathcal{M}^y=2$) mesh for the scalar,  which is temperature in the RB case.

\begin{figure}[htb!]
\begin{center}
 \includegraphics[width=0.49\textwidth]{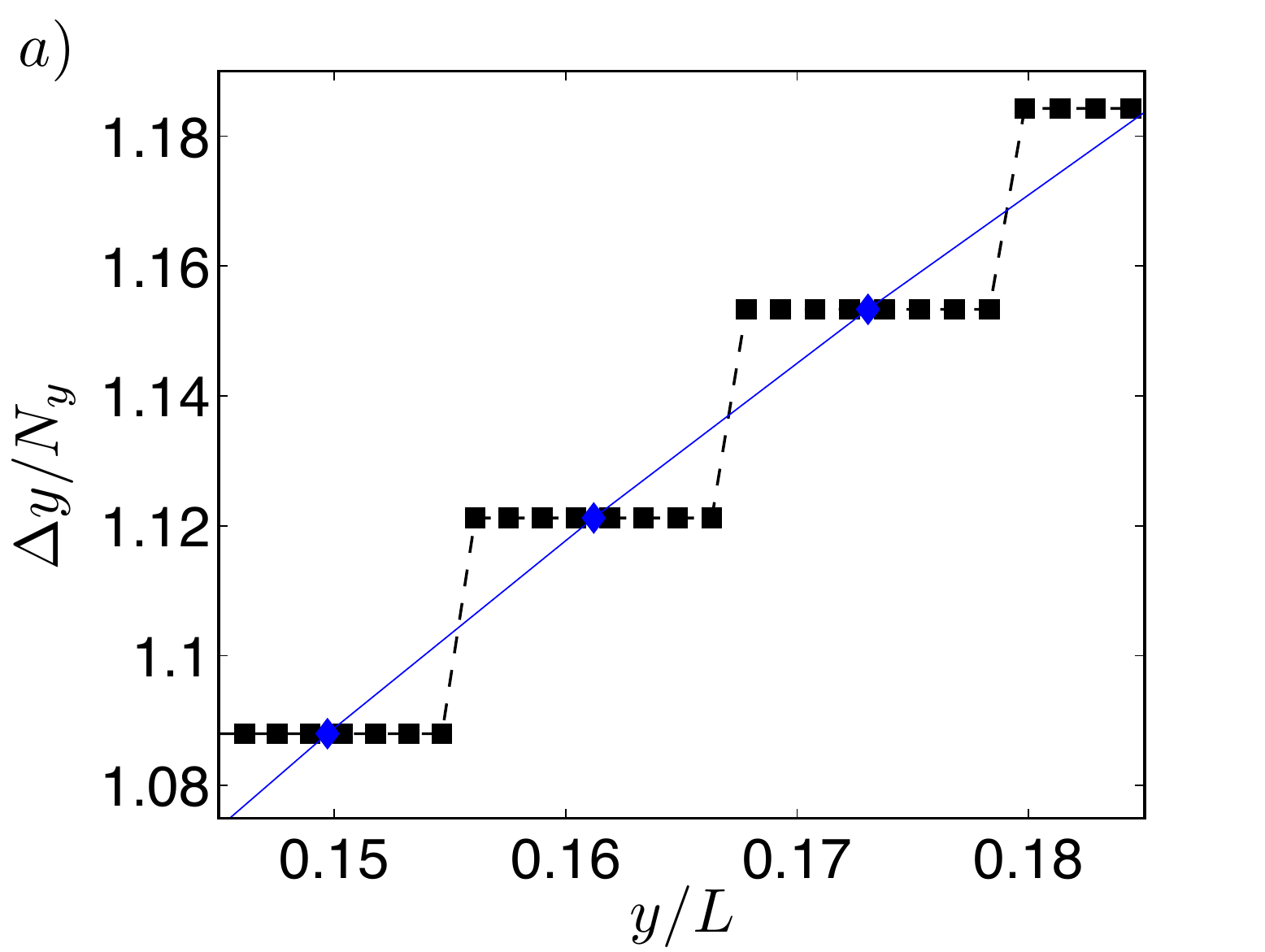}
 \includegraphics[width=0.49\textwidth]{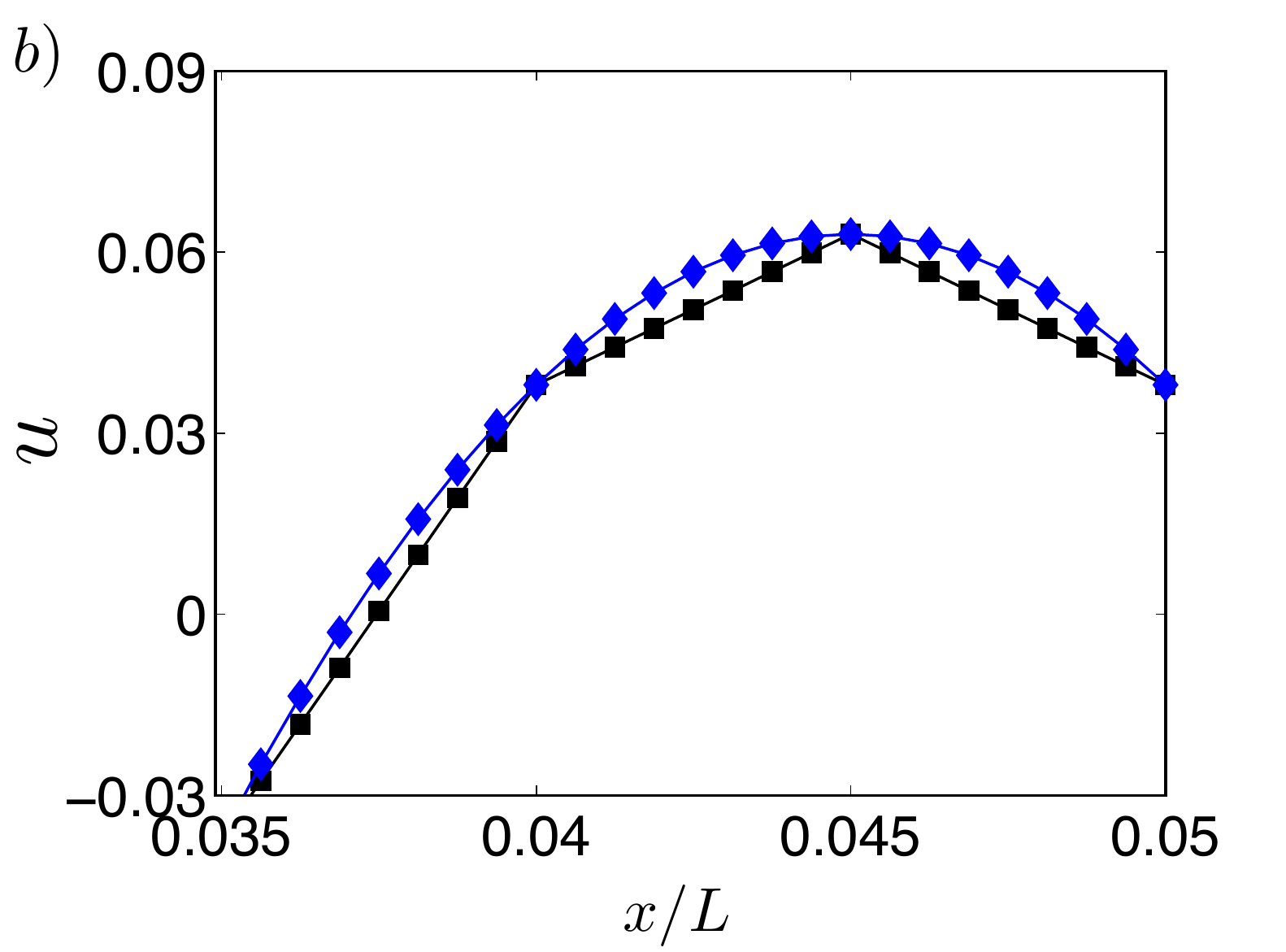}
\end{center}
\caption{({\it a}) Normalized metric $\Delta x/N_x$, where $N_x$ is the amount of grid points for a Chebychev-type grid clustering for a refined-mesh generated ($\mathcal{M}=8$) from a base mesh by splitting the base cells (squares), and the base mesh generated from the refined mesh (diamonds). The first method causes a staircase-like metric, which leads to spurious oscillations. ({\it b}) Comparison on interpolated velocity $v$ from a base mesh to the refined mesh ($\mathcal{M}=8$) using linear interpolation (squares) and cubic Hermite splines (diamonds). Both interpolations coincide at the base mesh points. The underlying basis for the interpolation is also plotted. }
\label{fig:metricvel}
\end{figure}

The method works as follows. We first generate the refined mesh over which the scalar field is discretized and then a coarser mesh is obtained by taking only one out of $\mathcal{M}^i$ nodes in the $i$-th direction. Note that when the mesh is uniform in space this is equivalent to start from the coarse cells and split them into $\mathcal{M}^i$ identical parts. For a non-uniform mesh, however, this naive splitting would result in a staircase distribution of the metrics for the fine grid with constant coefficients within each coarse element and with jumps across its boundaries. These discontinuities would locally decrease the accuracy of the discretization to first order and introduce spurious oscillations in the resolved fields. The difference between the two methods is shown in figure \ref{fig:metricvel}({\it a}) for a mesh obtained by a Chebychev--like distribution with 96 base nodes and a refinement factor of eight. In this chapter, most numerical examples are obtained using the same $\mathcal{M}^i$ in every direction therefore, unless otherwise specified, from here on we will use only $\mathcal{M}$ to indicate the isotropic refinement factor without specifying the direction. However, this is not required for the method to work and the same procedure can be applied to refinement levels different in each direction depending on the particular flow physics. An example will be shown in $\S$\ref{ssec:ddc}.

\begin{figure}[htb!]
\begin{center}
 \includegraphics[width=0.98\textwidth]{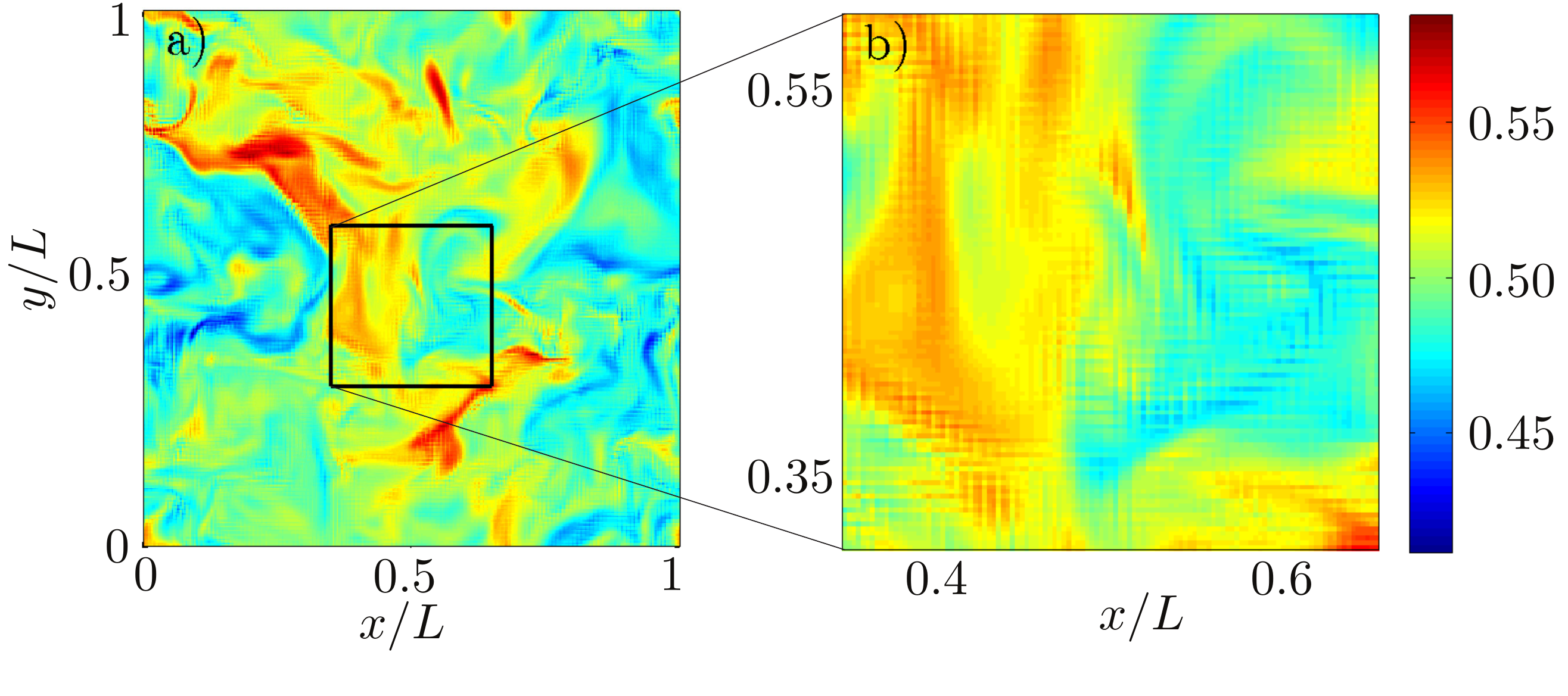}
\end{center}
\caption{({\it a}) Pseudocolour plot of temperature at the mid-height when using a refinement of $\mathcal{M}=3$ and linear interpolation for the velocities. Spurious oscillations every three points can be seen all over the domain. ({\it b}) Zoom-in of the region inside the black square of (\emph{a}). }
\label{fig:osc}
\end{figure}

In order to advect the scalar, a velocity field has to be projected from the base mesh onto the refined grid in a staggered arrangment with respect to the scalar. A first 
straightforward possibility consists of  using a tri--cubic Hermite spline interpolation, 
with a stencil of four points in every direction, for a total of sixtyfour points in 
three dimensions which, according to our numerical tests, is the minimum required. 
At the top and bottom boundaries, one--sided interpolation is used, which in principle 
is less accurate, but it is performed on a much finer grid, therefore the amplitude of 
the error is much smaller than in the bulk. The accuracy of Hermitian interpolation has 
proven to be sufficient in our turbulent flows, and it is comparable to that of B-splines 
\cite{hin13}. Hermitian interpolation, however, is preferred in this method as B--splines 
are much more computationally expensive.
Preliminary simulations have shown that a linear interpolation using a two point stencil,
though computationally cheaper, is not adequate since it results in a spatially 
interpolated velocity field which has equal gradients for all the refined points 
inside every base cell, and discontinuities at the base cell boundaries (figure \ref{fig:metricvel}({\it b})). 
This lack of homogeneity results in spurious oscillations in the scalar field, in particular around local maxima and minima of velocity. An example of these oscillations is shown
 in figure \ref{fig:osc} with the checkboard pattern given by the footprint of the base mesh.


The spatially interpolated velocity can then be used to advance equation (\ref{eq:scalareq}), 
and compute the values of the scalars at the new time. 
If a scalar couples back to the momentum field, like in the case of RB flow, a spatial filter
must be applied to calculate a ``coarsened'' scalar. In this case, an averaging using
equal weights within each refined cell is used. This averaged scalar is then used in equation (\ref{eq:navierstokes}) to advance momentum and pressure. Notice that in this case the scalar field is interpolated from a fine mesh onto a coarser one, therefore the previous problem of having equal fluxes between neighbouring cells is not encountered.

We stress that the interpolation of a velocity field between two different grids is a very dangerous operation since its effect is equivalent to that of a low--pass filter, which usually leads to loss of energy and generation or destruction of information. This is catastrophic in the DNS of turbulent flows where the dynamics are based on the energy cascade through the scales. It has even more serious consequences in a RB flow where the balance between thermal (potential) and kinetic energy determines the heat transfer. Nevertheless, if the base mesh already captures the smallest momentum structures, the field is smooth already at the scale of the base cell and an interpolation kernel that is continuous enough does not alter the energy content of the field. This has been directly verified with the DNS data by computing the relation (\ref{def2}) and the various meshes and comparing the results.

As mentioned above, interpolating the velocity field from the base to the refined mesh is the most immediate although it must be noted that it results in a \emph{non} solenoidal field. We wish to stress that this is not due to the tricubic 
Hermitian interpolation since, owing to the staggered arrangement of the velocity components, even a linearly interpolated velocity would be non solenoidal on the fine grid.

Indeed one may also just use the tricubic interpolation ``as-is''. Our simulations in this chapter indicate that the global responses and turbulent statistics show great agreement with those by using a single refined grid, and this non--solenoidal field has not resulted in apparent problems.  The residual divergence is in fact very small ($\sim{\cal O}(10^{-3})$) for most part of the flow domain as long as the base mesh is in the DNS range for the momentum. Higher residual divergence may be observed in some very localised regions which are usually close to the wall boundary, such as the places where the scaler plumes develop. Notice that at every time step the refined velocity field is interpolated from a solenoidal velocity field on the coarse grid, and the tricubic interpolation preserves the property of global divergence-free in our problems. Thus the error induced by the residual divergence should be small and does not accumulate during the time advance. Nevertheless, the effect may get worse as the grid becomes more non--uniform which could become a problem at higher drivings.

A drastic way of addressing this issue could be a ``minimum energy'' correction to 
fully remove the residual divergence of the interpolated velocity; this would require the
solution of a Poisson equation for a velocity correction, with the local divergence 
as source term. While the multiple resolutions strategy would remain favourable, solving 
a Poisson equation on the fine grid would negate one of the main advantages of the method. 

Alternatively, we propose here a locally mass--conserving interpolation which produces a divergence-free velocity field without solving the Poisson equation on the fine grid. Notice that the velocity field on the coarse grid is solenoidal in the sense of $\partial_x u+\partial_y v+\partial_z w = 0$ at the cell center. Thus we first compute $(\partial_x u, \partial_y v, \partial_z w)$ at the cell center on the coarse grid, then interpolate them to the cell center on the fine grid by using the tricubic interpolation. Since the three quantities are defined at the same locations on both the coarse and fine grids, the interpolation involves the same stencil points and we must have $\partial_x u+\partial_y v+\partial_z w = 0$ at the cell center on the fine grid. With the interpolated $(\partial_x u, \partial_y v, \partial_z w)$ the velocity field can be reconstructed. For example, one can conduct a two-dimensional tricubic interpolation for $u$ on an arbitrary $y$--$z$ plane, then integrate along the $x$--direction according the interpolated $\partial_x u$. $v$ can be treated similarly. For wall normal velocity $w$, one can just integrate $\partial_z w$ from the two plates where $w\equiv0$. 

Of course some price has to be paid by using this new interpolation. It requires more CPU time than the tricubic interpolation due to extra integration and data communication between processes. Another issue is the accumulated error of the integration. For instance, the integral of the interpolated $\partial_z w$ from one plate to the opposite one may not equal to zero, which will violate the impermeability condition at wall boundary. In practice, one can reduce the error by constructing $w$ from both plates and taking the average. In our simulation we found out that the error of $w$ at two plates is of order $10^{-6}$, and the total flux caused by this error is basically machine zero at both plates. Furthermore, the difference between the total kinetic energy of the coarse grid and fine grid is usually about $0.1\%$.

The previous discussion suggests that the proposed multiple resolution procedure can only work if the coarse mesh is fine enough to fully resolve the momentum field. In $\S$\ref{sec:res} numerical results will confirm this statement showing that when the base mesh adequately resolves the momentum field good results and CPU time saving can be obtained by refining only the grid for the scalar. On the other hand, if the coarse grid does not fully resolve the momentum field even very large values of $\mathcal{M}$ do not lead to correct results.

\subsection{Multiple resolution strategy in time}
\label{sec:mrest}

The multiple resolution in space 
entails that the node spacing for the scalars ($\Delta_S$) 
is smaller than that for the momentum ($\Delta_U$) and 
this has immediate consequences on the stability of the time integration because
of the explicit terms. Due to the
Courant--Friedichs--Lewy condition \cite{cou28}, $\Delta t\cdot {\rm max}[U/\Delta_S] \leq \mathcal{C}_{CFL}$, in fact, 
the time step must decrease by a factor
equal to the refinement $\mathcal{M}$ because ${\rm min}[ \Delta_S ] = {\rm min}[\Delta_U ]/\mathcal{M}$. 
As this small $\Delta t$  is not needed by the base mesh for momentum and pressure,
 the usage of the current approach becomes disadvantageous very rapidly, especially in high $Sc$ flows requiring high values of $\mathcal{M}$. However, a multiple resolution strategy can be applied also in time, 
by advancing the more expensive momentum and pressure with a larger time--step
and the scalar with a smaller one, using a temporally interpolated velocity.
In this way, the stability of the explicit terms in the scalar equation is retained
without the penalization of an unnecessary small integration step for all the 
other equations.

The integration of the scalar equation is therefore performed in ${\cal L}$ sub-steps, 
and at each intermediate $l$ time level the  velocity is linearly interpolated through 

\begin{equation}
 \bar{q}^l = \displaystyle\frac{{\cal L}-l}{{\cal L}}\bar{q}^n + \displaystyle\frac{l}{{\cal L}}\bar{q}^{n+1},
\end{equation}

\noindent where $\bar{q}^n$ is the spatially interpolated velocity at time step $n$. A simple linear interpolation is used, which is enough to ensure correctness as is demonstrated in $\S$\ref{sec:res}. This velocity is then used to advect the scalar(s) for every subtimestep using equation (\ref{eq:scalareq}) until the scalars have been advanced to the time $t^{n+1}$. Then, the velocity can be advanced a further time-step and the procedure repeated. If the maximum possible CFL is used for the velocity equations, which is usually the case, then ${\cal L}\ge \mathcal{M}$ must be satisfied  to ensure stability.

\section{Results and computational performance}
\label{sec:res}

In this section we will present the results of the multiple resolution method, and the associated saving in computational time and memory usage. A more extensive analysis will be performed in the first part for RB flow while, in the second part, additional results with anisotropic refinements will be shown for DDC flow.

\subsection{Rayleigh--B\'enard flow}
\label{ssec:rbf}

A series of rectangular RB simulations with aspect ratio $\Gamma=1$ were run to validate and to demonstrate the benefits of the
 described method.
Meshes of $96^2\times192$, $128^2\times256$ and $192^2\times384$ (only for $Pr=1$) were used for momentum, 
the $Ra$ was kept constant at $10^9$ while
$Pr$ was taken $Pr=1$ or $Pr=10$. In order to minimize the computational costs  
${\cal L}=\mathcal{M}$ was always used except for a specific set of runs in which the effects of ${\cal L}$
were isolated.

\begin{figure}[htb!]
\begin{center}
 \includegraphics[width=0.45\textwidth]{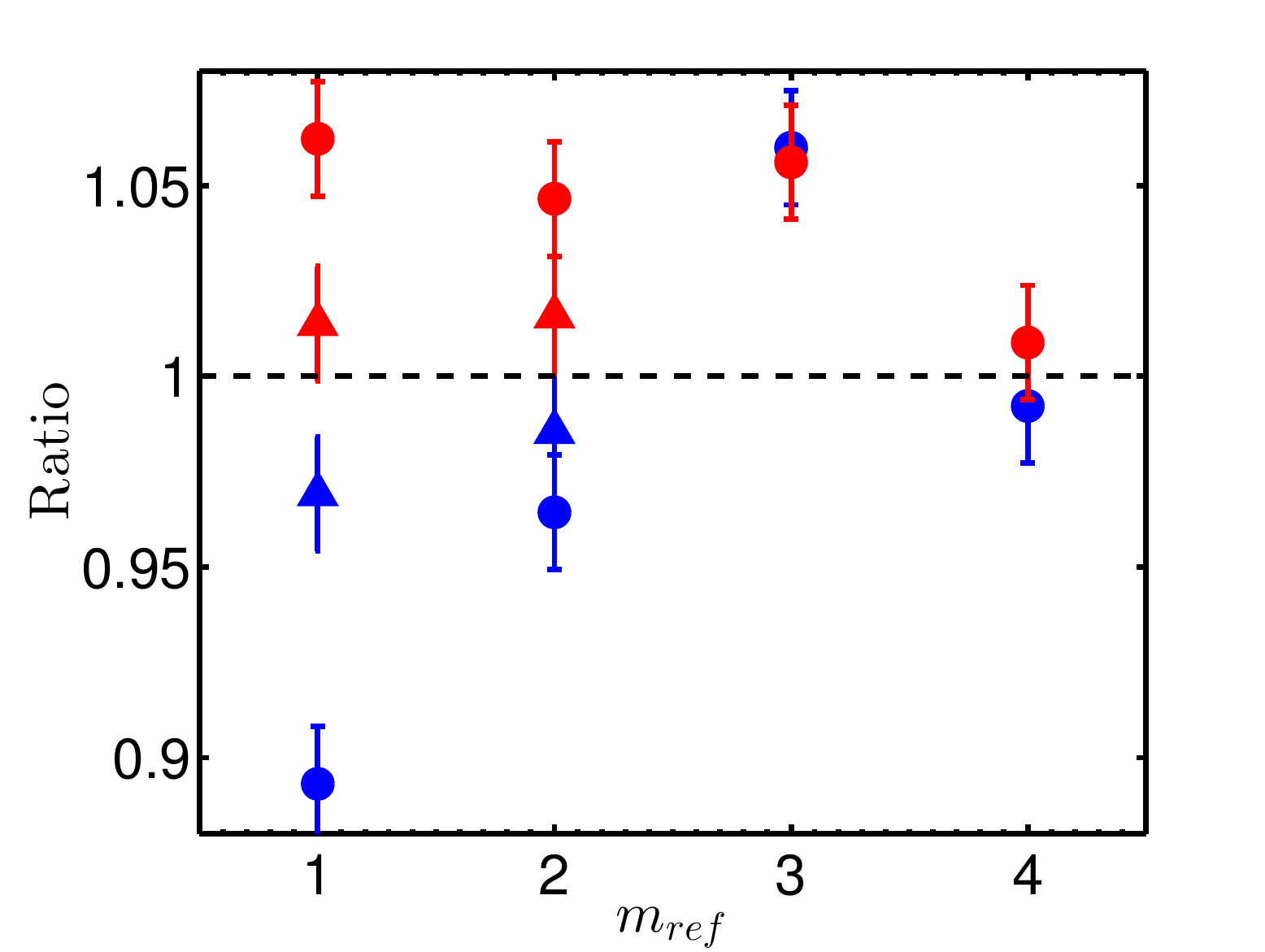} 
 \includegraphics[width=0.45\textwidth]{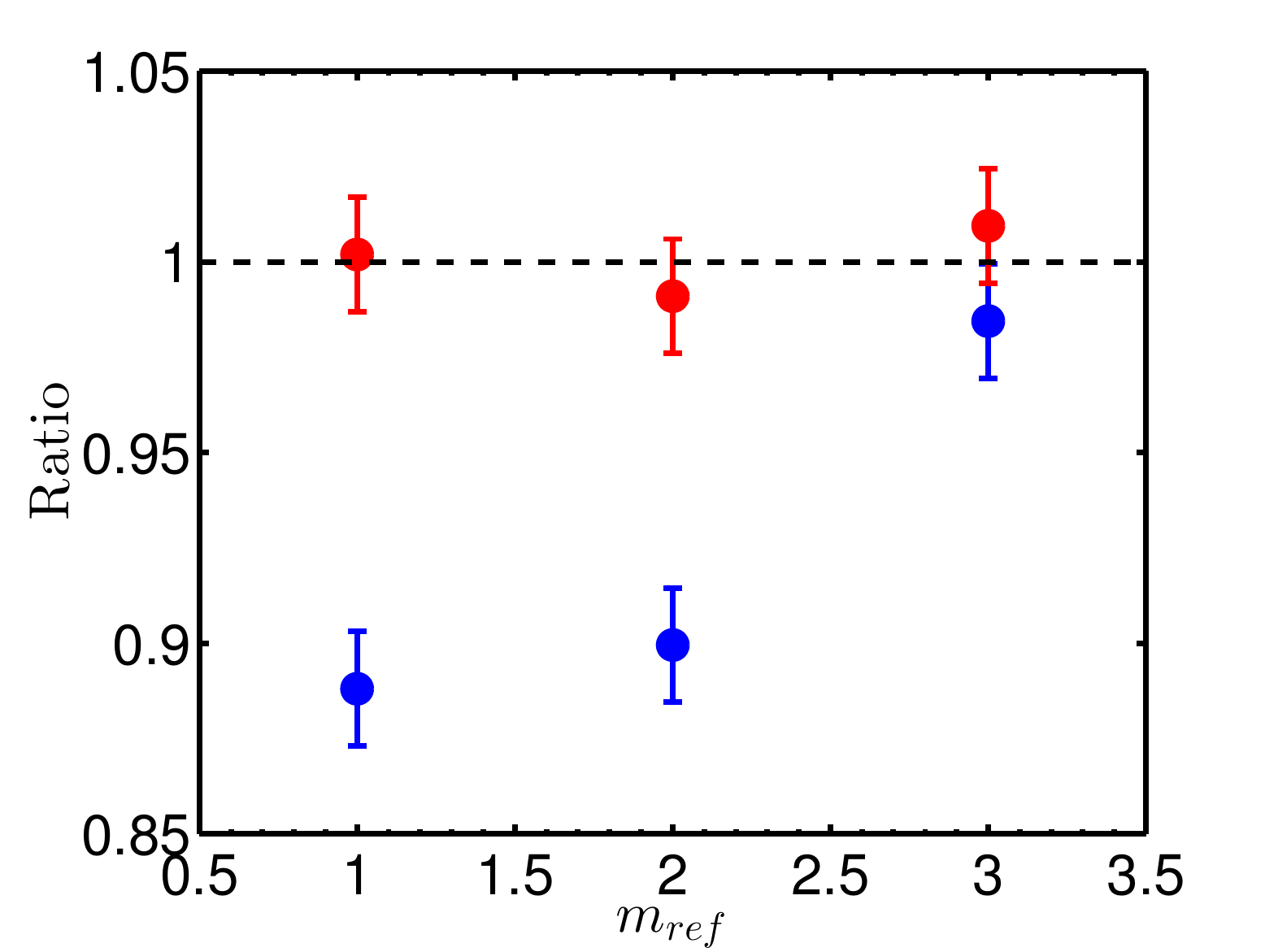} \\
 \includegraphics[width=0.45\textwidth]{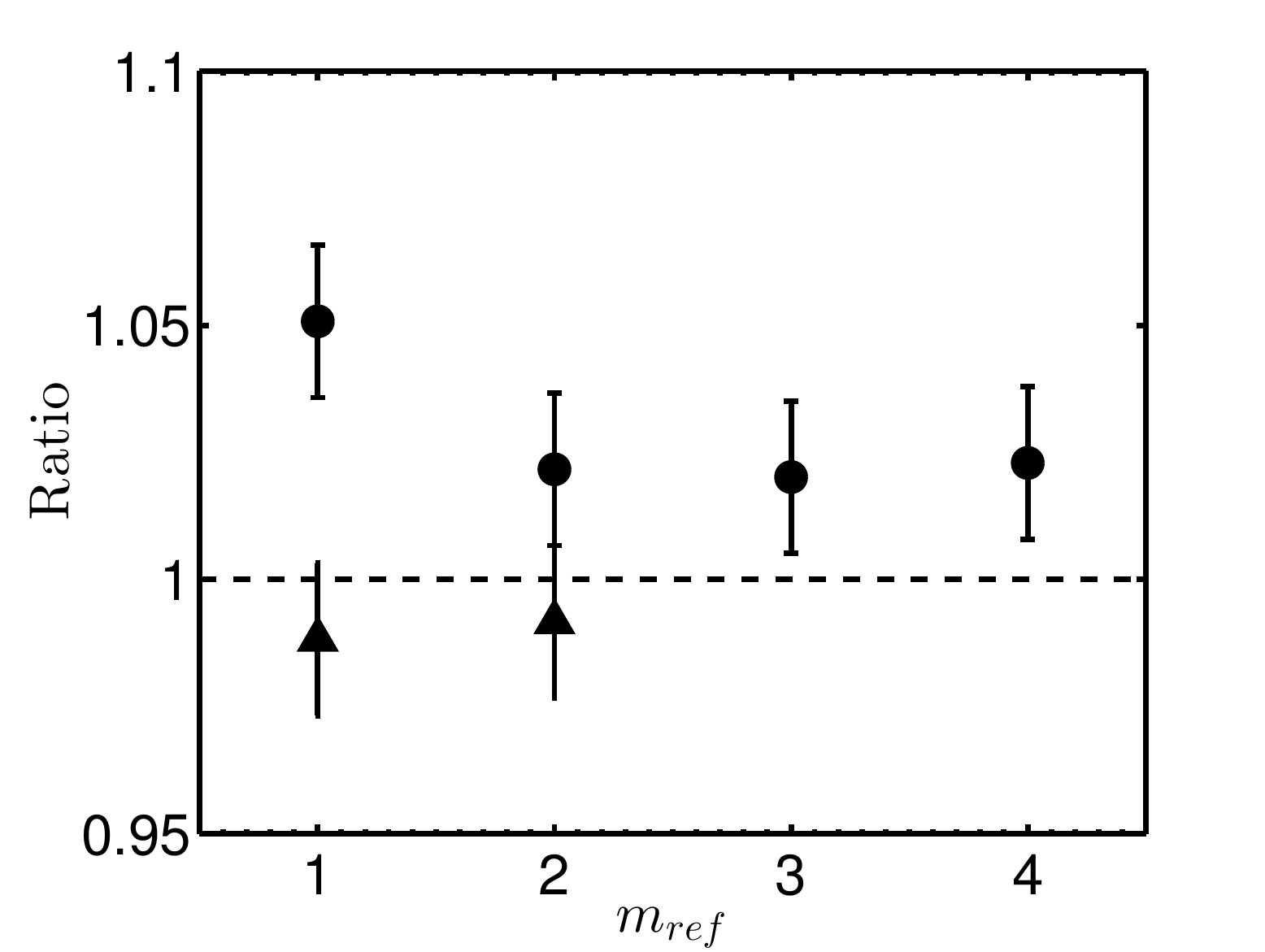}
 \includegraphics[width=0.45\textwidth]{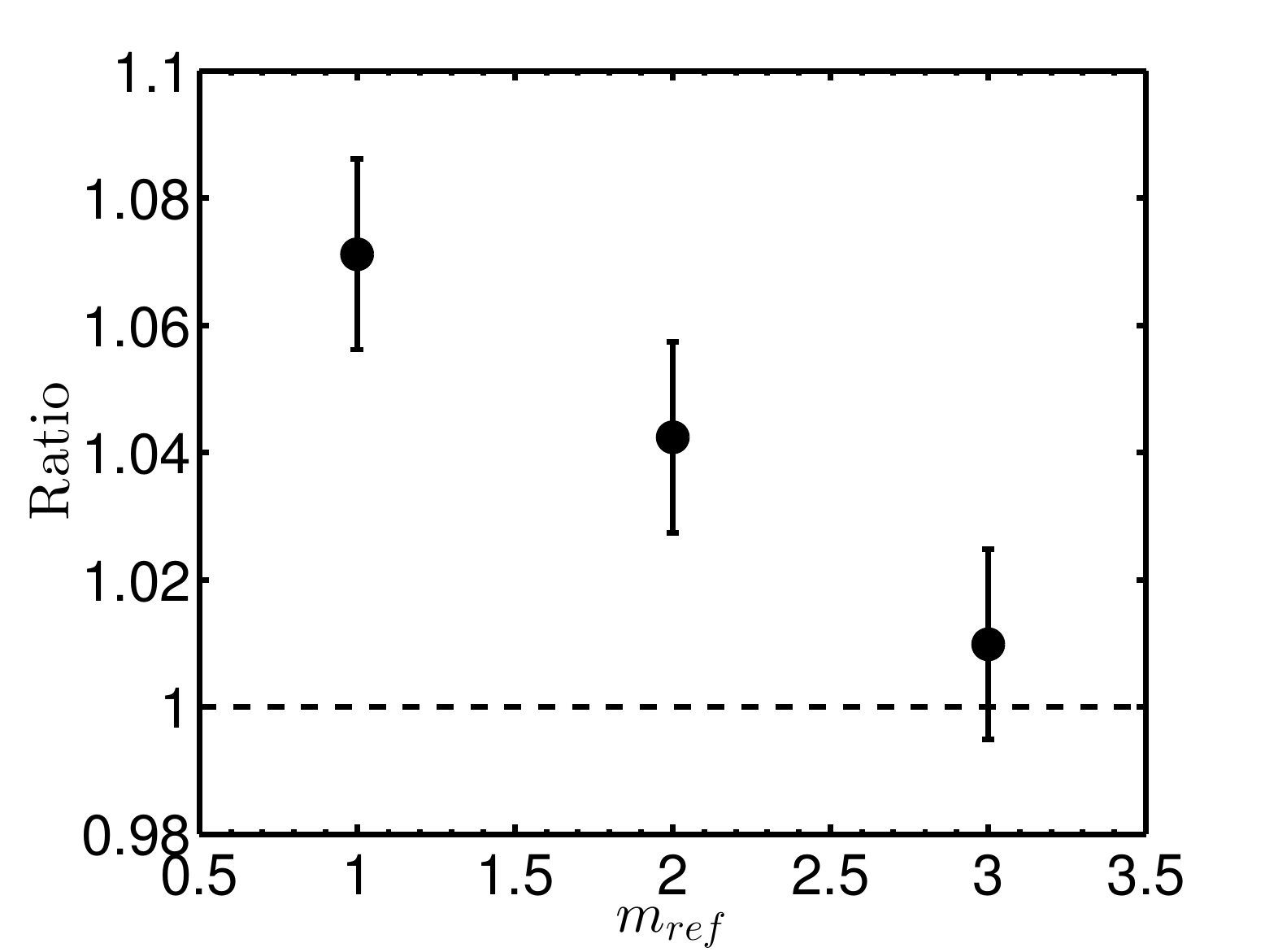}
\end{center}
\caption{({\it a}) Ratio between different ways of calculating $Nu$ against grid size for the two base meshes and increasing spatial refinement for $Pr=1$. ({\it b}) same as ({\it a}) for $Pr=10$. ({\it c}) convergence to an asymptotic value of $Nu_{u_z\theta}$ for increasing refinement and $Pr=1$. ({\it d}): same as ({\it c}) for $Pr=10$. Circles are for base meshes of $96^2\times192$, while triangles are for $192^2\times384$ base meshes. On the top panels, blue indicates the ratio $Nu_{\epsilon_\theta}/Nu_{u_z\theta}$ and red the ratio $Nu_{\epsilon_\nu}/Nu_{u_z\theta}$. }
\label{fig:nusmultres}
\end{figure}

Figure \ref{fig:nusmultres}(\emph{a},\emph{b}) shows the ratio between the different definitions
of $Nu$ and the maximum refinement level $\mathcal{M}$ while
the bottom panels report the convergence of $Nu_{u_z\theta}$ to the asymptotic reference value, calculated from a wiggle-free
simulation such as the one seen in Figure \ref{fig:osc}(\emph{b}).
The raw numerical values can be found in table \ref{tbl:multiple} at the end of the chapter.

As mentioned before, the multiple resolution strategy only works if the base mesh 
is sufficiently fine to fully resolve the momentum field. Looking at figure \ref{fig:nussingleres}(\emph{a}) we can see that at $Pr=1$ the $96^2\times192$ mesh is not sufficiently fine, and the different $Nu$ definitions do not show a monotonic convergence of the Nusselt ratios to the asymptote. It is interesting to note that for $\mathcal{M}=4$ the $Nu$ ratios get close to one, although the absolute values of $Nu$ are wrong and do not indicate a convergence towards the reference grid independent value. On the other hand, the $192^2\times384$ grid  yields a converged value of $Nu_{\epsilon_\nu}$ even though that resolution is not enough for the computation of $Nu_{\epsilon_\theta}$ that involves squared temperature gradients. For this case a refinement factor $\mathcal{M}={\cal L}=2$ for the temperature gives an appropriate resolution as shown by the Nusselt numbers that converge to the reference value. Although a factor two in space and time might seem to produce only limited benefits, we should consider that it implies a grid for the momentum and pressure with $2^3$ less elements than for the temperature. In addition, the momentum equations are solved only once every other time sub-step therefore, even if there is an overhead introduced by the interpolation of the fields, the CPU time savings are substantial. In our case the standard simulation on the single grid $384^2\times 768$ required for the integration of dimensionless time unit a wall-time of 36.4 minutes on 96 processors, for a total of 58.2 CPU hours using 13 GB of RAM memory. Using ${\cal L}=\mathcal{M}=2$, one dimensionless time unit required a wall-time of 36.7 minutes on 48 processors for a total of 29.4 CPU hours using 4.9 GB of RAM memory.

The method becomes even more advantageous as the Prandtl number increases. In fact, from equations 
(\ref{eq:navierstokes})--(\ref{eq:scalareq}) we see that the nondimensional diffusivities of
momentum and temperature, read $Re=\sqrt{Ra/Pr}$ and $Pe=\sqrt{RaPr}$,  respectively. Therefore for $Pr > 1$ the momentum field smoothens while the temperature field develops sharper gradients. This  results in larger Kolmogorov $\eta_K$ and smaller Batchelor $\eta_B$ scales that need different meshes to be properly resolved. Figure \ref{fig:nussingleres}(\emph{b}) shows that at $Pr=10$ the increased momentum diffusivity makes even the relatively coarse grid $128^2\times256$ adequate for the description of the momentum field. On the other hand, the same mesh is clearly too coarse for the scalar field as the ratio $Nu_{u_z\theta}/Nu_{\epsilon_\theta}$ strongly deviates from one. This grid, however, can be used to advance the momentum and to generate a refined mesh to advect the temperature. For this case the convergence of the Nusselt numbers to the reference value is obtained for $\mathcal{M}={\cal L} = 3$ that yields a computational gain around a factor $7$ and a reduction of RAM memory by a factor $3.5$ when compared to the reference cases using a single mesh.

Before concluding this section we point out that for all simulations we have used a refinement factor for the time step  ${\cal L}=\mathcal{M}$. Values of ${\cal L}$ smaller than $\mathcal{M}$ can be used provided the CFL number for momentum is reduced so that the scalar integration remains stable; this increases the number of time steps needed to advance the simulation over the same physical time resulting in an increased computational cost. On the other hand, further increasing ${\cal L}$ beyond $\mathcal{M}$ does not modify the results within statistical error and empirical evidence supporting this statement can be found in table \ref{tbl:multipletime} at the end of the chapter.

\subsection{Double diffusive convection}
\label{ssec:ddc}

\begin{figure}[htb!]
\begin{center}
 \includegraphics[width=0.95\textwidth]{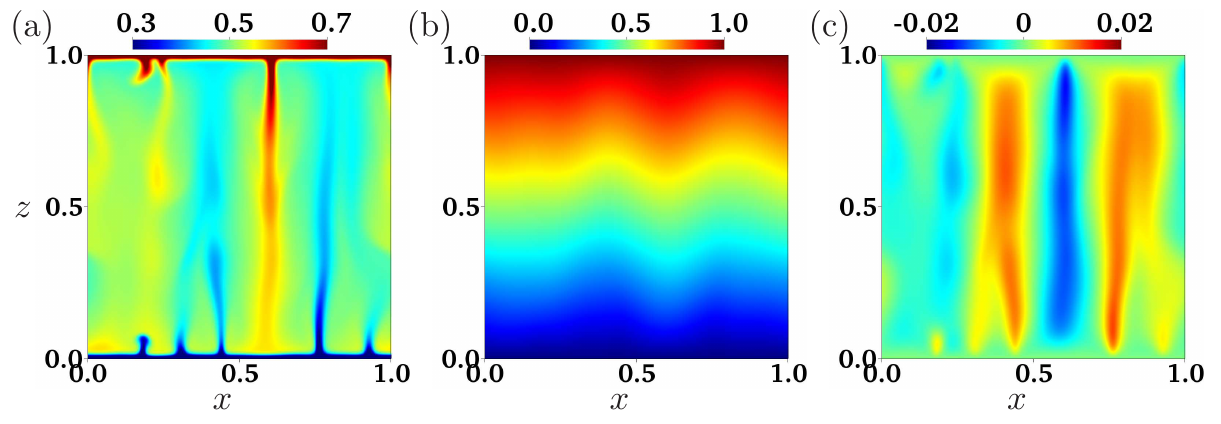}
\end{center}
\caption{Instantaneous snapshot of ({\it a}) salinity ($Sc_s=700$), ({\it b}) temperature ($Pr=7$) and ({\it c}) vertical velocity in a DDC simulation at drivings of $Ra_s=5\times10^7$ and $Ra_T=5\times10^5$. This results in $Le=100$ and $R_\rho=1$.}
\label{fig:ddcvis}
\end{figure}

Double--diffusive convection (DDC) is a system where two scalars with very different diffusivities are coupled to the flow field, one of which is stabilizing and the other destabilizing. A relevant example of this system is the ocean, where the scalars are temperature ($Pr\approx7$) and salinity ($Sc\approx700$). The former being warmer at the top surface and cooler at the bottom stabilizes the flow while the latter has the opposite effect because a higher salinity at the  top boundary results in a denser, sinking fluid. A snapshot of the flow in a DDC system for a geometry similar to RB can be seen in figure \ref{fig:ddcvis}. Very sharp gradients of salinity can be observed, while the temperature field is in a quasi--diffusive state.

In this case, the computational gains associated with a multiple resolution can be even larger than before, owing to the very large $Sc$ number of salt in water. In the simulations of Ref. \cite{yan15}, the temperature is discretized without further refinement, for salinity a factor $\mathcal{M}=3$ is used for all three directions. In this simulation we did not use the maximum possible CFL for the velocity equations, and ${\cal L} =1$ for the time step.

Similar to the RB case, the results in terms of heat transfer, salinity fluxes and turbulence statistics agree within the uncertainty of $2$--$3\%$ with those obtained using a single refined grid for all the variables. Nevertheless, using the present numerical approach momentum and temperature equations are solved on a mesh with $3^3$ less nodes than that for salinity, i.e., the coarse grid of $144\times144\times144$ and the refined grid of $432\times432\times432$. By using two 12-core 2.6 GHz Intel Xeon E5-2690v3 (Haswell) CPUs with 12 MPI process and 2 OPENMP threads, the CPU time for one time step is about 12.9 seconds for using single refined grid, 2.69 seconds for multi grids with tricubic interpolation, and 2.98 seconds for multi grids with our mass conserving interpolation, yielding a ssaving factor that exceeds $4$. A decrease of RAM memory usage by more than $50\%$ is also achieved by using the multi resolution strategy.

Before concluding this section it is worthwhile to discuss briefly why in DDC it is possible to simulate a flow with a Schmidt number of $700$ without using a refinement factor $\sqrt{Sc} \simeq 27$. The DDC equations in nondimensional form read
\begin{eqnarray}
  \partial_t u_i + u_j \partial_j u_i &=& -\partial_i p + 
  \sqrt{\frac{Pr_S}{Ra_S}}\,\partial_{jj} u_i + (R_\rho \theta - S) \delta_{iz}, \\
  \partial_t \theta + u_j\partial_j \theta &=& Le \sqrt{\frac{1}{Pr_S Ra_S}}\,\partial_{jj}\theta, \\
  \partial_t S + u_j\partial_j S &=& \sqrt{\frac{1}{Pr_S Ra_S}}\,\partial_{jj} S.
\label{eq:ddc}
\end{eqnarray}
where the flow parameters are defined as 
$Ra_T = g \beta_T \Delta T L /(\nu \kappa_T)$, $Pr_T = \nu/\kappa_T$,
$Ra_S = g \beta_S \Delta S L /(\nu \kappa_S)$ and $Pr_S = \nu/\kappa_S$, $Le=Pr_S/Pr_T$ and $R_\rho=Ra_T Le/Ra_S$.

It can be noted that the diffusivity of momentum $Re = \sqrt{Pr_S/Ra_S}$ increases with
$Pr_S$. Therefore for large enough values the flow does not fully transition to turbulence.
On the other hand, the equations for salinity and temperature are linear and they can not
sustain the cascade without a turbulent velocity field. This is indeed the case for the flow
parameters of the present numerical example (see also Ref. \cite{yan15}) where
the fully turbulent three dimensional cascade cannot be achieved 
and a factor $\mathcal{M}= \sqrt{Sc}\simeq 27$ is not required for salinity. Even so, the multiple
resolutions strategy results in a substantial gain factor in computational time and 
RAM memory occupation when compared against the single grid strategy.

\section{Summary and conclusions}
\label{sec:conc}

In this paper we have presented a numerical strategy for the direct numerical
simulation of turbulent flows 
with active and/or passive scalar fields without over-resolving
the momentum equation
and its pressure correction. This is certainly the case of flows with scalar diffusivity
smaller than the kinematic viscosity ($Pr$ or $Sc>1$). Substantial
computational time and memory occupation savings are even obtained for equally diffusive fields with Schmidt numbers of order unity. The different requirements for spatial discretization of scalars with respect to momentum
originate not only from the diffusivity but also from the pressure. Its non--local effect
was found to smoothen the momentum gradients 
and thereby reduce the resolution requirements with respect to resolving the scalar field. This scenario modifies the picture obtained from dimensional analysis that compares only the Kolmogorov and the Batchelor scales.

To reduce computational costs, a multiple resolutions strategy was developed in
which momentum is discretized
on a base mesh while scalars are discretized on a refined mesh. To solve the scalar 
diffusion--advection equation, momentum is spatially interpolated onto the refined grid 
and either tricubic Hermitian splines or a more sophisticated procedure, based on the
interpolation of velocity divergence, are proposed. 
The scalar is advanced in time, and if necessary, 
coarsened to couple it back to the momentum equations. Due to stability constraints on the non--linear terms, the scalar is advanced in time using a refined 
timestep. Velocity is linearly interpolated in time for all the intermediate timesteps. The optimal amount
of substeps ${\cal L}$ coincides with the grid refinement factor $\mathcal{M}$, when it is isotropic, or with
${\rm max}[\mathcal{M}^i]$ when it is anisotropic.

The method was applied to Rayleigh-B\'enard convection, and decoupling the grid resolutions was found to result
in computational speedups around two for Prandtl number unity, and seven
 for $Pr=10$.
This strategy was also applied to high $Sc$ flows, also resulting in computational advantages
of approximately four.
Due to the large costs, both in operations, memory usage and in communication associated to solving the Poisson equation,
we expect the gains to increase for larger grids and larger drivings. This is because the Poisson solver is the most expensive part of advancing the Navier--Stokes equations in time, and this does not scale linearly with the amount of points, while the scalar diffusion-advection equations do. We expect gains of about three to four times for RB simulations at $Pr=1$ and $Ra=10^{12}$ with production grids of about 1 billion points and $\mathcal{M}=2$, planned for the future. Also- the memory consumption is heavily reduced, by a factor 2.6x with a refinement of two, and this makes some simulations possible on supercomputers with a lower memory per core and decreases the dependence on high CPU--memory bandwidth.

Once again,
it is crucial that the base mesh is fine enough to correctly resolve the momentum field. Adding more refinement to the scalar mesh when the velocity grid is insufficient does not give an improvement of the quality of the results, and it might even lead to the suppression of small scales that violate energy conservation. This method could in principle be additionally applied to flows with $Pr<1$. Obviously, in this case the velocity field should be solved on a mesh finer than that of the scalar. Although explicit numerical tests have not been attempted, we expect that the  computational overhead introduced by the interpolation and coarsening of the fields overcomes the advantages produced by solving the scalar equation on a coarser mesh.

\textit{Acknowledgments:} We acknowledge FOM, an ERC Advanced Grant and computing time from SurfSARA (granted through NWO) and PRACE grant 2013091966.

\appendix

\section{Numerical details}
\label{sec:app}

\begin{table}[hp]
  \begin{center}
  \begin{tabular}{|c|c|c|c|c|c|}
  \hline
  $N_x\times N_y\times N_z$ &  $Pr$ & $Nu_{u_z\theta}$ & $Nu_{\theta_w}$ & $Nu_{\epsilon_\nu}$ & $Nu_{\epsilon_\theta}$ \\ 
  \hline
   \hline
  $96 \times96 \times192$ &  $1$  & $66.8$ & $67.2$ & $71.0$ & $59.7$ \\
  $128\times128\times256$ &  $1$  & $64.9$ & $64.6$ & $67.1$ & $60.5$ \\
  $192\times192\times384$ &  $1$  & $63.2$ & $63.2$ & $64.1$ & $61.3$ \\
  $256\times256\times512$ &  $1$  & $63.8$ & $63.6$ & $64.6$ & $62.5$ \\
  $384\times384\times768$ &  $1$  & $64.0$ & $63.6$ & $64.3$ & $63.2$ \\
  \hline
  $96 \times96 \times192$ &  $10$ & $68.4$ & $67.9$ & $68.6$ & $60.8$ \\
  $128\times128\times256$ &  $10$ & $65.5$ & $65.2$ & $65.7$ & $61.0$ \\
  $192\times192\times384$ &  $10$ & $63.2$ & $63.0$ & $63.1$ & $61.0$ \\
  $256\times256\times512$ &  $10$ & $63.6$ & $63.9$ & $63.6$ & $62.4$ \\
  $384\times384\times768$ &  $10$ & $63.4$ & $63.9$ & $63.4$ & $63.3$ \\
  \hline
 \end{tabular}
 \caption{Details of grid resolution used for standard single grid runs. Simulations were run until $Nu_{u_z\theta}$ achieved
 1\% temporal convergence. All the simulations are performed at $Ra=10^9$ and $\Gamma=1$. 
The first column shows resolution, the second $Pr$, while the 
 other four show the results of the different definitions of $Nu$.}
 \label{tbl:single}
\end{center}
\end{table}

\begin{table}[hp]
  \begin{center}
  \begin{tabular}{|c|c|c|c|c|c|c|c|}
  \hline
  $N_x\times N_y\times N_z$ & $\mathcal{M}$ &  $Pr$ & $Nu_{u_z\theta}$ & $Nu_{\theta_w}$ & $Nu_{\epsilon_\nu}$ & $Nu_{\epsilon_\theta}$ \\ 
  \hline
  \hline
  $96 \times96 \times192$ & $2$ & $1$ & $65.5$ & $65.4$ & $68.4$ & $63.0$ \\
  $96 \times96 \times192$ & $3$ & $1$ & $65.6$ & $65.3$ & $69.3$ & $69.5$ \\
  $96 \times96 \times192$ & $4$ & $1$ & $65.4$ & $65.4$ & $66.0$ & $64.9$ \\
  $128\times128\times256$ & $2$ & $1$ & $64.4$ & $64.4$ & $66.7$ & $66.7$ \\
  $128\times128\times256$ & $3$ & $1$ & $64.4$ & $64.4$ & $66.9$ & $66.9$ \\
  $192\times192\times384$ & $2$ & $1$ & $63.5$ & $63.4$ & $62.6$ & $64.5$ \\
  \hline
  $96 \times96 \times192$ & $2$ & $10$  & $66.6$ & $64.8$ & $60.0$ & $66.0$ \\
  $96 \times96 \times192$ & $3$ & $10$  & $64.5$ & $64.5$ & $63.5$ & $63.5$ \\
  $96 \times96 \times192$ & $4$ & $10$  & $65.4$ & $65.0$ & $64.2$ & $64.2$ \\
  $128\times128\times256$ & $2$ & $10$  & $64.3$ & $64.2$ & $64.7$ & $63.0$ \\
  $128\times128\times256$ & $3$ & $10$  & $64.6$ & $64.1$ & $62.2$ & $64.8$ \\
  \hline
 \end{tabular}
 \caption{Details of grid resolution used for multiple resolutions runs. Simulations were run until $Nu_{u_z\theta}$ achieved  1\% temporal convergence. 
All the simulations are performed at $Ra=10^9$ and $\Gamma=1$.
The first column shows resolution, the second shows the refinement
 of the scalar grid in all directions, the third $Pr$, while the 
 other four show the results for the different definitions of $Nu$. For all simulations 
${\cal L}=\mathcal{M}$. }
 \label{tbl:multiple}
\end{center}
\end{table}

\begin{table}[hp]
  \begin{center}
  \begin{tabular}{|c|c|c|c|c|c|}
  \hline
   ${\cal L}$ & $\mathcal{C}_{CFL}$ & $Nu_{u_z\theta}$ & $Nu_{\theta_w}$ & $Nu_{\epsilon_\nu}$ & $Nu_{\epsilon_\theta}$ \\ 
  \hline
  \hline
   $1$ & $0.6$ & $64.5$ & $64.6$ & $67.3$ & $64.0$ \\
   $2$ & $1.2$ & $64.4$ & $64.4$ & $66.8$ & $63.5$ \\
   $3$ & $1.2$ & $64.6$ & $64.3$ & $67.2$ & $63.5$ \\
  \hline
 \end{tabular}
 \caption{Details of the testing for the temporal multiple
 resolutions. Simulations were run until $Nu_{u_z\theta}$ achieved
 1\% temporal convergence. All the simulations are performed at $Ra=10^9$ and $\Gamma=1$ on
a grid $128\times128\times256$ with $\mathcal{M}=2$.
The first column shows the time refinement level ${\cal L}$, the second 
the maximum CFL computed on the momentum grid, while the 
 last four show the results of the different definitions of $Nu$. }
 \label{tbl:multipletime}
\end{center}
\end{table}



\bibliographystyle{model1-num-names}
\bibliography{/home/rodosti/NetworkHomeCopy/literatur}







\end{document}